# Learn to Adapt: Robust Drift Detection in Security Domain

Aditya Kuppa*, Nhien-An Le-Khac*

*School of Computer Science, University College Dublin, Ireland*

**Abstract**

Deploying robust machine learning models has to account for concept drifts arising due to the dynamically changing and non-stationary nature of data. Addressing drifts is particularly imperative in the security domain due to the ever-evolving threat landscape and lack of sufficiently labeled training data at the deployment time leading to performance degradation.

Recently proposed concept drift detection methods in literature tackle this problem by identifying the changes in feature/data distributions and periodically retraining the models to learn new concepts. While these types of strategies should absolutely be conducted when possible, they are not robust towards attacker-induced drifts and suffer from a delay in detecting new attacks. We aim to address these shortcomings in this work. we propose a robust drift detector that not only identifies drifted samples but also discovers new classes as they arrive in an on-line fashion.

We evaluate the proposed method with two security-relevant data sets – network intrusion data set released in 2018 and APT Command and Control dataset combined with web categorization data. Our evaluation shows that our drifting detection method is not only highly accurate but also robust towards adversarial drifts and discovers new classes from drifted samples.

## 1. Introduction

Machine learning (ML) and its variants are widely used for a variety of tasks in cybersecurity from malware classification, intrusion detection, online abuse detection analysis to malicious event detection [1]. ML models deployed in production are prone to *concept drifts*, a phenomenon which is characterized by the continuous changes in the latent data generating function (e.g., class distributions) over time or simply from having a lack of access to truly stable and well-defined collection of instances [2] at the training time.

*Corresponding author
 *Email addresses:* aditya.kuppa@ucdconnect.ie (Aditya Kuppa ), an.lekhac@ucd.ie (Nhien-An Le-Khac )

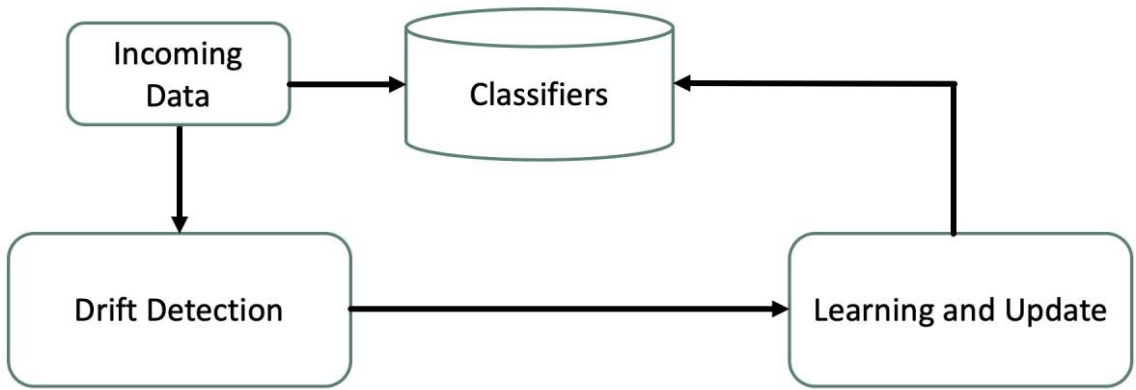

Figure 1: Drift Detection System consist of three modules – (1) The *Classifiers* which actually gets trained on historic data. (2) *Drift Detection module* processes the incoming data and identifies the drifted samples and updates the learning algorithm when required. (3) A *Learning and update module* defines the dissimilarity measure to understand any type of drifts before being presented to the learning algorithm. It also tracks the performance of the learning algorithm and updates the classifier periodically.

For example, in image and text domain the models accuracy dropped from 3% − 15% on CIFAR-10 and 11% − 14% on ImageNet when tested on drifted distributions of the original and new test sets [3]. Similar phenomenon has been observed on question-answering data [5] and other variants of ImageNet [4].

More recently, each application domain has seen a shift from training ML models from an offline to an online mode, where new data is collected, labeled, and models are trained in a continuous fashion to adapt to changing environments with research focused primarily to address concept drifts. There are several works on concept drift detection/adaptation [6, 7], most of them focus on drifts that are non-adversarial in nature and assume that (a) Incoming new data is of high quality, with little to no noise; (b) Drift direction, type, and scale are known in advance; (c) Immediate and proportional feedback available to perform model correction; and (d) Measure the performance degradation of drift detectors[21] by employing ensemble, windowing, and statistical techniques to deal with drifts whose characteristics are not known at training time.

These combinations of assumptions describe a closed dataset where data distribution parameters are known *apriori* and the extensive subject matter expertise is available to adjust distribution parameters of models when drift occurs. Also, in an adversarial setting, one has to consider the potential presence of a malicious attacker, aiming at compromising the underlying drift detector. Adversarial learning in the presence of malicious data and poisoning attacks recently gained a significant attention [5]. However, most of the research focuses on dealing with attacking the classifiers at train/test time by introducing corrupted/outliers/adversarial samples [10, 9] and detecting the potentially dangerous samples [8] with a little focus on attacker-induced drifts.

Accounting for concept drifts is very important in the cybersecurity domain. Models and insights are likely to become obsolete quickly due to the ever-changing threat landscape. Data volumes change due to technological advancement, new protocols and software appear at a rapid pace, and malicious activity also follows trends and can be temporal in nature. Also, drifts observed in cybersecurity domain are unique in nature due to (a) absence of timely feedback due to scale of data - manually labeling of even 0.01% of web data (>1M samples per day) will require analysts to work continuously for 8 hours; (b) drifts can be unknown and unbounded, due random shifts in user/attacker behavior; and (c) new data may contain significant amounts of noise and may be irrelevant to the problem at hand. Existing methods in the literature address the concept drift by designing them as adaptive systems [2]. These systems consist of three modules as illustrated in Figure 1

One can view training a model in the presence of concept drift as a case of *incremental learning* where the model learns one example at a time. Periodically retraining the model with fresh data, adjusting the shifts in feature/data by re-weighting the samples inversely proportional to their age [11] or extracting feature representations, which are invariant [12] and, employing active learning techniques [13, 20, 19] are some of the common techniques employed in literature to address the concept drifts. In the cybersecurity domain, previous works employed techniques such as checking the prediction confidence score of incoming samples and using a threshold [14] function to decide if the sample is drifted from training distribution. Recent works employ the distance function to measure the dissimilarity between a new sample and trained classes to establish the sample is drifted [17, 15, 16]. Studies which design robust drift detectors in the presence of adversarial drifts and identifying new concepts (classes) simultaneously in an automated fashion are rare in literature and we aim to bridge the gap in this work.

In this paper, we present a novel drift detection method that not only identifies drifted samples from the originally trained distribution but also immune to adversary-induced malicious drifts in an automated fashion. First, we map the high-dimensional input vector to low-dimensional output space using an embedding network with the contrastive loss that brings similar samples closer to each other and pushes away dissimilar samples far from each other. Next, a Nearest Class Mean (NCM) classifier is trained for a drift detection task, with a distance metric that captures both instance level (euclidean) and group level (geometric) fidelity to make the detector robust towards adversarial drifts. Finally, to discover new classes in the drifted samples, we leverage a recursive update mechanism by accumulating drifted vectors and create a new class when a pre-defined threshold is crossed. We evaluate the proposed method with two security-relevant data sets – network intrusion data set released in 2018 [36] and APT Command and Control dataset combined with web categorization data. Our evaluation shows that our drifting detection method is not only highly accurate but also robust towards adversarial drifts and discovers new classes from drifted samples.

In summary, this paper has three main contributions:

- We propose a novel drift detection method to address concept drift based on the contrastive loss objective and Nearest Class Mean (NCM) classifier to train a robust drift detector that is immune to attacker-induced drifts and effective in identifying drifted/new samples/classes.

- We also devise a recursive update mechanism to discover new classes which are not available at training time from drifted samples in an automated fashion.

- We evaluate the proposed method with two security-relevant datasets and adversarial evaluation settings and show that the proposed method is effective in drift detection, identifying new classes, and robust towards attacker-induced drifts.

## 2. Preliminaries

### 2.1. Concept Drift

Concept drift [2, 24] refers to the condition when the test time data/feature distributions deviate from the original training time distributions and in turn causes a shift in the true decision boundary. It is mainly observed in non-stationary and dynamic environments where latent data generating function changes over time or lack of access to highly stable and well-defined data samples at training time [2].

In a typical classification task, a predictor $\phi$ is learned on labeled data $\{(X_i, Y_i)\}_{i=1}^{n}$ where $X_i \in X$, $Y_i \in Y = \{1, \ldots, K\}$ denote the features (covarites), and the labels respectively, and the pairs $(X_i, Y_i)$, $i = 1, \ldots, n$ are sampled in i.i.d (Independent and Identically Distributed ). from some unknown joint distribution $P$ over $X \times Y$. Concept drift at a given time $t$, is measured in terms of change in joint probability function $P_t(X, y) = P_t(X) \times P_t(y|X)$, which in turn depends on both components $P_t(X)$ and $P_t(y|X)$. Drifts, which affect only $P_t(X)$ component i.e. features/labels and not changing the underlying conditional distribution $P_t(y|X)$ are categorized into covariate/label drifts. Drifts, which affect $P_t(y|X)$ change the decision boundaries of the classifier and can lead to degradation of performance. Generally, concept drift focuses on changes to any/both $P_t(y|X)$ and $P_t(X)$ components, as each one captures important dynamics of the learning environment.

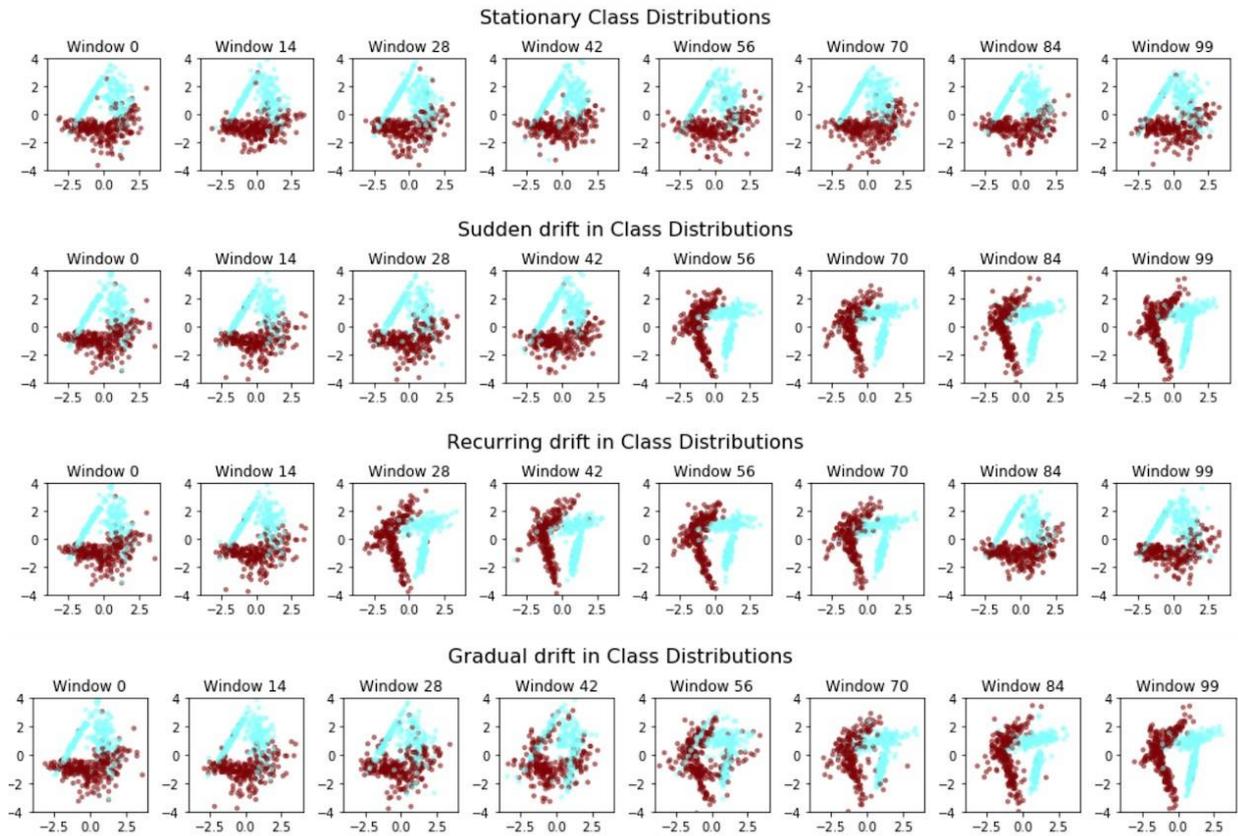

Figure 2: Illustration of different types of naturally occurring concept drifts using a toy dataset[1].

To understand concept drifts one has to take into account the drift characteristics like (a) **Influence on decision boundaries** [18] – Unlike feature/label drifts, which have no influence on the learned decision rules or classification boundaries, concept drifts may affect the performance of the underlying classifier changing the posterior probabilities and impacting unconditional probability density functions. Drifts which change the posterior probabilities $P(y|\mathbf{x})$ are identified as *real* concept drifts and *virtual* drifts occur when the distribution of features $\mathbf{x}$ change over time i.e. these affect the unconditional probability distribution $P(\mathbf{x})$. Both can occur separately or simultaneously and may have different impacts on the classifier's performance; (b) **Locality** – Drifts can be local or global [22] depending on the changes affecting large data vs small regions of feature space or set of instances or a subset of classes. Depending on the locality of change one has to retrain the whole classifier or parts of model or sub-models, leading to a more efficient adaptation; (c) **Recurrence** – Observations leading to drifts may repeat over time and can be sudden, gradual or sporadic [23]. Figure 2.1 illustrates different types of naturally occurring drifts in a toy dataset. *Sudden drift* results due to an abrupt change in Instance distribution. *Incremental drift* occurs when the distribution of the old concept is decreasing and the new concept keeps increasing over time. *Gradual drift* is identified when two incoming distributions are observed during the duration of the drift, with the old concept appearing with decreasing frequency. Recurring drifts are handled by training dedicated models to identify previously seen drifted patterns. (d) **Feature Decay and Noise** – As new behaviors appear in data with time, new features may emerge expanding the feature space and, old features may become irrelevant to the learning task at hand [24]. Similarly, corruption in the feature values or class labels in the form of noise may give a false sense of drifts in the data distribution [25].

In a multi-class classification setting, drifts can occur due to multiple reasons: (a) **Class Evolution** – Samples belonging to classes that are not part of training data or show some significant change in behavior of existing class fall under out/in-class drifts. For example, a new class of malware discovered in the wild is an out-class drifted sample and a variant of known malware with small changes can be categorized as an in-class drifting sample. In/out class drifts can be addressed by retraining the underlying classifier with new training samples, but in order to successfully re-train the underlying drift detector has to identify and flag the samples as drifted or classify as unknown/new class. (b) **Adversarial drifts:** Not all drifts are natural, samples that belong to an existing class or a new class can be injected by a malicious adversary to fool the underlying drift detector. In this case, the original classifier can make mistakes depending on the threat model of the adversary. For example, adversaries can induce feature/label drifts by flipping labels or with modified feature values targeting $P_t(X)$ component by instance-based poisoning attacks. Attacks, which focus on changing the class boundaries by class(concept) based poisoning attacks aim to modify the $P_t(y|X)$ component of the concept drift. These attacks can significantly impair the adaptation to the actual concept drift and may misguide the underlying drift detector to adapt to changes that are not natural.

*Contrastive learning.* Contrastive learning was initially proposed to learn the high quality representation in a self-supervised manner and has been applied in various fields with great success[26]. At its core, contrastive learning is motivated by InfoMax principle [27], which aims in maximizing the mutual information (MI) between two samples/views/distributions[28]. Unlike other sample-wise loss functions (softmax, hinge, or mean squared error) the contrastive loss is discriminatory in nature and measures the dis/similarities between the sample pairs, which results in separating dense samples in deep latent space by preserving manifold locality and maximizing the manifold margin. For our drift detection problem, we rarely have access to all classes at training time and have to learn a model using labeled data from the training environment and unlabeled data from the deployment environment. This is a common scenario in real-world ML applications in many domains. Ideally, we need an objective function that can learn tight class boundaries with no *supervision* at training time, and give a measure of class membership at deployment time. Given the properties of the contrastive loss function, we employ it for drift detection problem.

*Embedding Networks.* Embedding networks map high dimensional data into low dimensional outputs by performing the automatic feature extraction and separating the samples by their similarity/dissimilarity that is represented by some distance metrics [30]. Having trained an embedding network we can use the embedding space for downstream tasks such as classification, clustering, etc. Here class boundaries are learned by some loss objective and class representative points in the embedding space are called *prototype* of class [29]. In our work, we use a combination of mean square error and contrastive loss to train the embedding network.

## 3. Motivating example

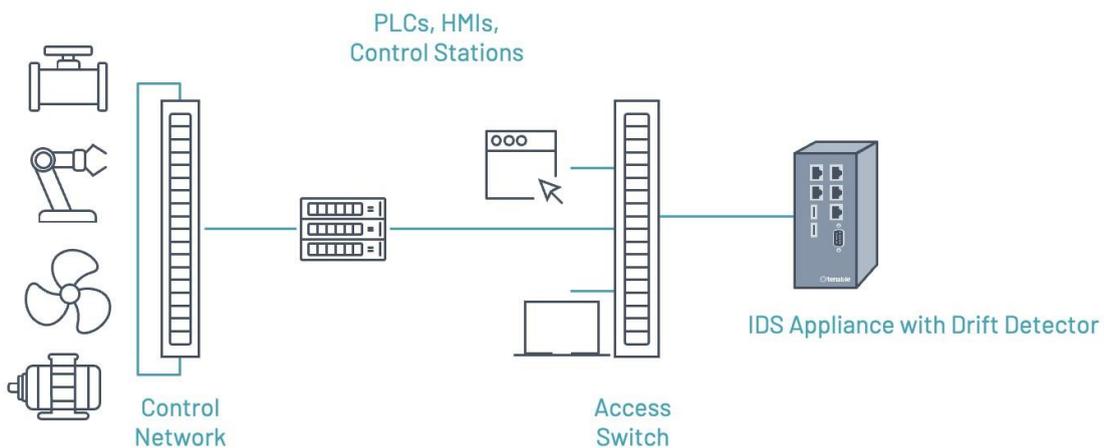

Figure 3: Industrial Control Systems deployment

In this section, we cover our main motivation towards designing a robust drift detector for real life usecase. Let's take an example of an intrusion detection system (IDS) to detect cyber attacks on industrial control systems(ICS). Figure 3 illustrates the typical real-world deployment of an IDS system for ICS. The sensors are distributed and the readings are sent to Programmable Logic Controller(PLC)/Human Machine Interface(HMI) for further analysis over a monitored network. The IDS system is retrained periodically with new data to accommodate new traffic patterns that emerge over time. A drift detector module is used to identify the data points which deviate from training time distributions and these data points are used to re-train the IDS periodically to address the overall change in system behavior. The goal of IDS and drift detectors is to protect the ICS from false/manipulated/malicious inputs sent by attackers and adapt the system to new changes.

The IDS/drift detector may use a learning system to discover the malicious traffic. For a user and attacker alike, the underlying learning system is a black box. In order to evade the deployed system, the attacker has to first gain control of a subset of sensors and manipulate the sensor readings by injecting malicious code/commands slowly over a period of time in the direction that would raise no alarms in both IDS and drift detectors and make sure the manipulated data is recorded as normal. Attacks of this nature have been observed in the real life wherein an attacker has gained access to the control panel of the water treatment plant and tried to change the values of the sensor to dangerous levels [2].

Previous works in the literature have addressed the problem by assuming non-stationarity of concept/label – The number of concepts (e.g. classes, environment, label, contexts) evolves over time. All these works tackle the problem in a non-adversarial setting. In our work, we aim to fill this gap by creating a robust detector and testing methods to improve the robustness of the underlying drift detectors in an adversarial setting.

*3.1. Threat Models*

The robustness property of ML systems is defined in terms of threat model assumptions. Several attempts have been made to categorize threat models for ML systems [31, 32]. Threat model assumptions for a system can be defined in terms of attacker *knowledge, capability, and goal* – targetted vs non-targeted, black/white/grey/no-box knowledge of model architecture, feature extraction functions, training/test data distributions, etc and attacker goals are categorized based on security vs privacy compromises of the underlying system. Attack *strategy* – how/where the attack can be executed for example at the training phase vs inference phase. Also, an attack can be carried out by training a surrogate model for the query reduction, parameter extraction, membership inference, and model stealing, etc. Attack *characteristics* describe the properties of each attack step such as iterative vs one-shot, perturbation size, and satisfy the domain constraints vs feature space attack.

In our threat model attacker has two evasion goals – (a) to evade the underlying drift detector which learns in an online fashion i.e. re-training itself as new samples arrive over time, by injecting instance-based drifts (poisoning samples) at different time windows so that the drifting samples will be misclassified as non-drift samples. (b) to evade the classifier the attacker injects concept drifts by poisoning class distributions of known classes so that the injected samples are classified as non-malicious.

---

[2]https://www.dragos.com/blog/industry-news/recommendations-following-the-oldsmar-water-treatment-facility-cyber-attack/

To reflect the real world attack scenario, we assume the attacker has no knowledge of classifier architecture, retraining intervals, defense methods but has access to training/test data distribution, inference API, feature transformation functions, and feature set. The attacker is capable of manipulating training data and can execute the attack in an iterative fashion.

## 4. Proposed model

Let the inputs are defined as $X_1, X_2, ...$, where $X_t = \{x_{t1}, x_{t2}, ..., x_{tN}\}$ is a set of incoming samples and $Y_t = \{y_{t1}, y_{t2}, ..., y_{tN}\}$ is a set of classes present in $X_t$, where N denotes the number of samples in each time window $t$. The model is trained at time $t_1$ and we have only access to data of $X^t, Y^t$ and do not have access to all classes $K$. Our task is to identify drifted samples in next time windows $t_2, t_3, ...$ in a sequential fashion and discover new classes as new data samples arrive. We assume the data distribution is locally stationary, i.e. stationary within a local time window $t$.

*Design.* The main function of a robust drift detector is to identify how similar/dissimilar the incoming samples are with the original training set and also be immune to attacker-induced adversarial drifts. To achieve this objective we need a distance function that supports *instance-level* fidelity i.e. encouraging the samples of the same class to be as close as possible while pushing samples from different classes to larger distance in some latent space. Similarly, to address adversarial drifts the objective function has to be aware of *group-level* fidelity i.e. for a new incoming sample distribution - the distribution distances between all the samples of the same class have to be minimum from the new input distribution. The intuition of distance function comes from the properties of adversarial subspaces: (a) compared to data manifold, they are found in low probability; (b) class distributions generally vary from their closest data sub-manifold; (c) adversarial samples are found closer to nearest neighbour proximity to the unperturbed sample than to any other neighbour in the training or test set. To apply the distance function efficiently on the high-dimensional input vector, we use an embedding network that performs automatic feature extraction and maps the input vector to low-dimensional latent output vectors. Now, we can leverage the learned embedding network to apply a distance function, to determine the similarity/dissimilarity between the training distributions and new incoming samples. This distance measure can help us to identify the drifted samples. In summary, we need a loss function for feature extraction and dimension reduction to learn the embedding network and a robust distance function that can identify the drift samples.

*4.1. Drift Detection System*

For training an embedding network, we use an autoencoder, which is trained on Mean Square Error (MSE) reconstruction loss (Equation 1) and contrastive loss (Equation 2) to capture the class level distinction in embedding space.

$$L_{MSE} = \|g(h(x)) - x\|_2^2 \tag{1}$$

where $h(\cdot)$ and $g(\cdot)$ denote the encoder and the decoder operation, respectively. The central idea in contrastive loss is to bring similar samples closer and push away dissimilar samples far from each other. One way to achieve this is to use a similarity metric that measures the closeness between the embeddings of two samples. The contrastive loss tries to make the positive pairs attracted and the negative samples separated, i.e., the positive alignment and negative separation.(2)

$$\mathcal{L}_{contrastive} = -\sum_i^B \log\left[\frac{e^{s(h_i, v_k)}}{\sum_j^P e^{s(h_i, v_j)}}\right] \tag{2}$$

$$s(h_i, v_j) = \frac{f_\theta(x_i) \cdot v_j}{\|f_\theta(x_i)\|\|v_j\|} \cdot \frac{1}{\tau}$$

Here, given an instance, $x_i$, that belongs to a particular class, $k$, we encourage its representation, $h_i = f_\theta(x_i)$, to be similar to the same class, $v_k$, and dissimilar to the remaining classes, $v_j, j \neq k$. We quantify this similarity, $s(h_i, v_k)$, by using the cosine similarity with a temperature parameter, $\tau$ in a batch $B$. The intuition is that each record, in being attracted to a diverse set of representations that belong to the same class and should become invariant to intra-class differences. Once the embedding network is trained, we can use the class prototypes to train the nearest class mean (NCM) classifier [33] to identify drifted samples. NCM is a distance-based classifier that assigns an incoming sample to the class with the closest mean and the success of NCM classifiers critically depends on the distance metric. In our problem we aim to achieve both instance and group level fidelity, so choosing a distance function that captures both instance level (euclidean) and group level (geometric) helps the detector more robust to adversarial drifts.

$$D = d_R(\boldsymbol{h}_j, \boldsymbol{h}_c) + \lambda_1(d_E(\boldsymbol{h}_j, \boldsymbol{h}_c)) \tag{3}$$

where $\boldsymbol{h}_c$ is the *prototype* of class $c$, $d_E$ is the euclidean distance between the class centriod of and incoming point in latent space, $d_R$ is Riemannian divergence (geometric distance) and, $\lambda_1 > 0$ is a tuning parameter.

The NCM is trained on class prototypes $h_c$ and mean of training samples available at a given time.

$$c_j^* = \underset{c \in C}{\mathrm{argmin}}\, D(\mathbf{h}_j, \mathbf{h}_c) \tag{4}$$

$$\mathbf{h}_c = \frac{1}{n_c} \sum_i [y_i = c]\, \mathbf{h}_i \tag{5}$$

where $c_j^*$ gives the class prototype for a given class, and $n_c$ is the number of training samples seen for class $c$.

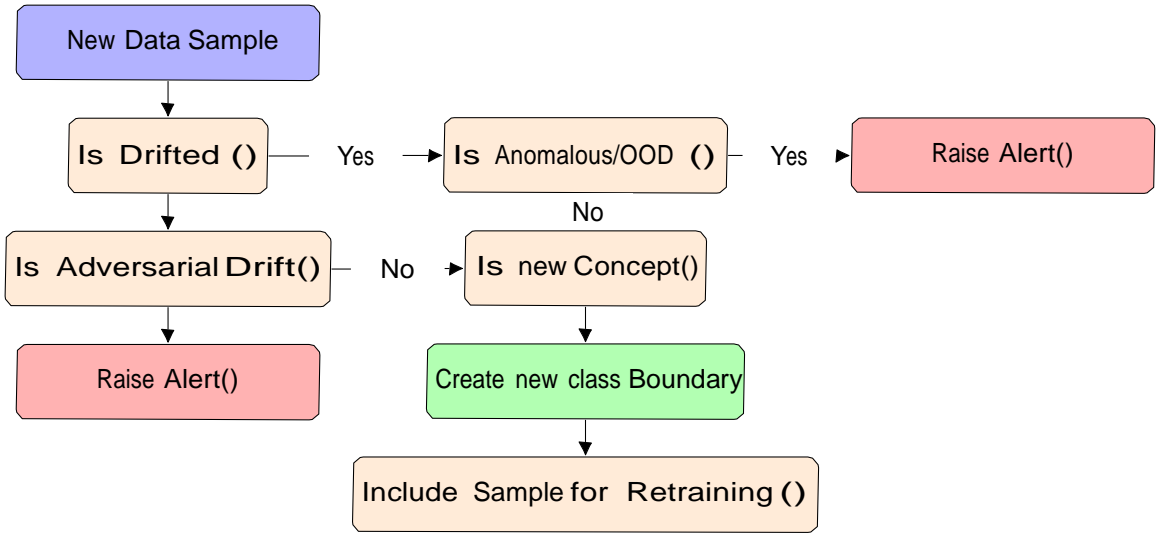

Figure 4: Flow chart of the proposed method – When a new sample arrives, it is examined for its anomalous/drifted/OOD properties. Depending on the characteristics of the sample a alert is raised or included as new training sample

*4.2. Identifying new Concepts*

At time window $t$ we have only access to data of $X_t, Y_t$ and we cannot compute the prototype means $h$ of all classes at $t_{i+1}$ as the class prototypes are not known. As new data samples arrive over time, the drift detector can tag samples as drifted with respect to classes in known *training* data. To discover new concepts in the drifted samples, we need an update mechanism that not only discovers new class prototypes from drifted samples but also refreshes the previously computed prototypes with new classes in an automated fashion. To address this problem we leverage the embedding vectors of all the drifted samples $\mathbf{z}^t_{drift}$, which are generated from encoder network.

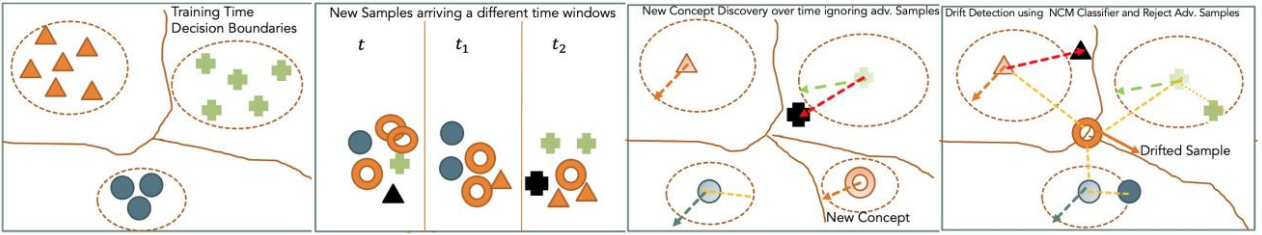

Figure 5: Illustration of the proposed drift detection method. From Left to Right: (a) Tight decision boundaries are learned for three classes at training time. (b) Data arriving at 3 different time intervals with known, unknown classes and, adversarial samples (samples in black) (c) Drift sample identification by calculating mean distances of samples from class prototypes and NCM classifier. Distance metric also helps in rejecting the adv.samples (d) New concept discovered with clear class boundaries as new samples arrive over time .

$$\mathbf{z}_{drift}^{t \to t+1} = h_c^{t+1} \quad h_c^t = \triangle_{drift}^{t-1 \to t} + \delta^{t-1 \to t}$$
$$\delta_i^{t-1 \to t} = \mathbf{z}_{i\,drift}^t - \mathbf{z}_{i\,drift}^{t-1} \quad y_i \in c^t, \tag{6}$$

We recursively accumulate the drifted vectors $\delta$ till it crosses a pre-defined threshold $T$ (NMC mean of original training classes) and create new mean prototypes $h_{new,i}$ as shown in Equation 6. Then encoder network is re-trained with new training data with the same loss function as described above. In summary, as shown in Fig. 5, we monitor the concept drift of the incoming data during the $t_1$ with respect to $t$ time window. This results in recursive accumulation of drift vectors $\mathbf{z}^t_{drift}$ that is used to form new class prototypes (classes). Figure 4 illustrates the decision flow in the proposed system.

*Adversarial Drifts.* For adversarial drifts, the attacker injects adversarial samples either in the training dataset $X_t$ or data specific to a set of time windows $X_{t_n}$ such that the underlying drift detector and/or classifier misses the detection. In particular, the attacker can carefully *modify*, *delete*, and/or *insert* some training examples in $X_t$ and classifier $\phi$, $\phi(X_t, \mathbf{x}) = \phi(X_t', \mathbf{x})$ for many testing examples $\mathbf{x}$ or some attacker-chosen $\mathbf{x}$, where $X_t'$ is the drifted dataset. We note that modifying a training example means modifying its feature vector and/or label. For injecting instance-based drifts (poisoning samples) we use a well-known ensemble learning method bootstrap aggregating (Bagging) [34] to divide dataset $X_t$ into subsamples with replacement and inject $\mathbf{L}_{inst}$ in each sub-samples. The adversarial examples are generated by projected gradient descent (PGD) method [41] to flip their class labels and use the bagging approach described above to inject samples in different batches. This will ensure the drift samples are distributed in each window/batch of the data and also across the population. For concept drifts injection attacks where the class distributions have to be poisoned, we use MAA [35] on the dataset $X_t$ and divide it into various data distributions $z_1, z_2..z_i$. For a successful attack, one has to induce minimal distortions on the input distribution $z_a$ to move the decision boundary to $z_i$ where $a$ is the natural sample distribution and $i$ is the target sample distribution. Perturbations based on data distribution helps us to satisfy the data manifold constraint i.e. adversarial sample $x_{injected}$ needs to be in the same distributions to fool the underlying detector. To summarize, the adversary aims to poison the data distributions in some lower dimension(embedding) space so that any downstream task using this embedding will be classified incorrectly.

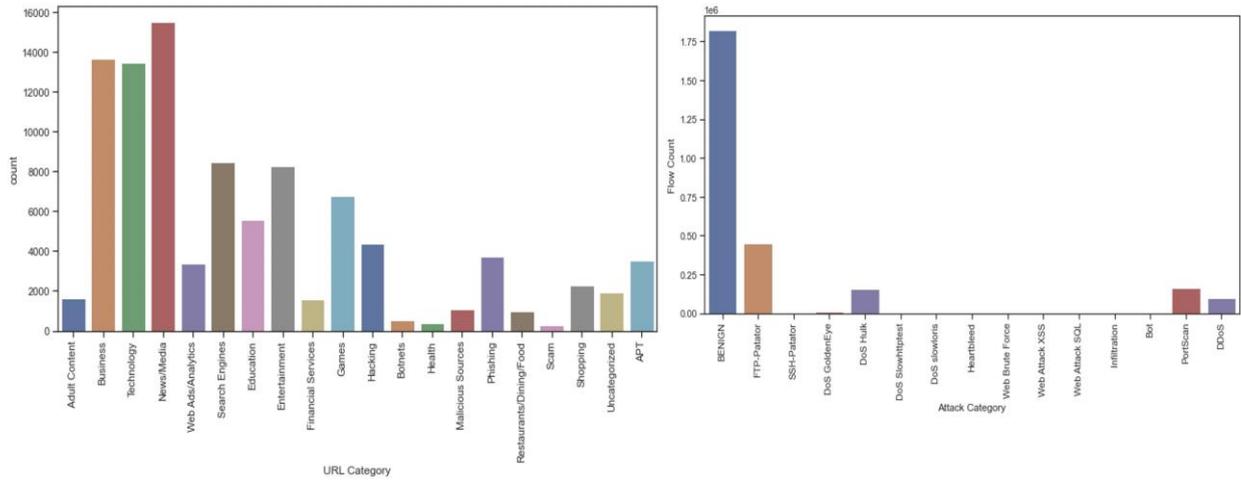

Figure 6: Dataset distributions used in this study

## 5. Experiment Setup

We choose two different security-relevant systems for our evaluation. Each system has a main classifier, which is trained on the main task and the underlying drift detector for identifying concept drifts in the data as illustrated in Figure 1.

*Network DataSet.* One of the major applications of anomaly detection in the field of cybersecurity is Intrusion Detection System (IDS) [36]. Unfortunately, most of the publicly available datasets lack diversity of the traffic, volumes, and real-world attacks [36]. The dataset covers seven different attack scenarios: Brute-force, Heartbleed, Botnet, Denial of Service (DoS), Distributed DoS (DDoS), web attacks, and infiltration of the network from inside. The infrastructure used to generate attack and benign traffic reflects a real-world setting with 420 victim machines, 30 servers, and 50 attack machines. Recently Engelen et al. [37] identified a set of errors in the original data set and released a modified version that addresses the flaws in data collection, feature extraction implementation, and attack labeling process. We use the modified version of *CICFlowMeter-V3* to extract 72 features from the captured traffic.

Features are extracted from bidirectional flows. Statistical time-related features are calculated separately for both directions. TCP flows are terminated by FIN packet and UDP flows are terminated by a flow timeout, which is set to 600 seconds. There are 8 groups of features that are extracted from raw pcaps: (a) Forward Inter Arrival Time, the time between two packets sent forward direction (mean, min,max, std); (b) Backward Inter Arrival Time, the time between two packets sent backwards (mean, min, max, std); (c) Flow Inter Arrival Time, the time between two packets sent in either direction (mean, min, max, std); (d) Active-Idle Time, amount of time flow was idle before becoming active (mean, min, max, std) and amount of time flow was active before becoming active (mean, min, max, std); (e) Flags based features – Number of times the URG, PSH flags are set both forward and backward direction; (f) Flow characteristics – bytes per second, packets per second, length of flow (mean,min,max,std) and download and upload ratio of bytes; (g) Packet count with flags – FIN, SYN, RST, PUSH, ACK, URG, CWE and ECE; (h) Average number of bytes and packets sent in forward and backward direction in the initial window, bulk rate, and sub flows.

*APT C&C and Web Categorisation DataSet.* Advanced Persistent Threats (APTs) campaigns are more challenging to detect than regular malware campaigns. APT operators tend to be well-trained in cybersecurity coming from military and governmental organizations, academia, and R&D entities [38]. Hawk-Eye [38] dataset is curated data of APT domains used for Command and Control (C&C) infrastructure. It consists of 3894 FQDNs and 1894 domains with metadata specific to APT campaign ID, the period of activity, the report source, and other relevant data such as the IP addresses of C&C servers. 175 total features are extracted, which are categorized into semantic, contextual, hybrid and holistic features.

Web page filtering is used in enterprises to enforce and update corporate security policies across the network. Given a URL, the web categorization engine (WCE) assigns a category to these sites, based on the categories returned granular control is enforced inside the enterprises network. New and unknown web content is sent to WEC for real-time categorization. We sample randomly 100,000 web URLs from a private dataset and use this dataset in our experiments. 134 Lexical and host-based features are extracted from the dataset such as count of special characters in URL and each part of URL, TLD in arguments, file extension, characters count, the ratio of digits to its length, the ratio of the number of consonants, and vowels to URL length, entropy, whois registration details, name servers, and hosting/infrastructure information. We combine the APT C&C dataset with the web pages categorization dataset. This dataset represents real-life imbalanced data. The feature size is limited to 134 in the combined dataset as not all features in each dataset intersected. Figure 6 summarizes various attack categories found in both the data set used in our study.

*Evaluation Metrics.* We use three evaluation metrics, each one for measuring the performance of the concept drift detection module, new concept discovery module and, robustness towards adversarial drifts. For concept drift detection, we randomly pick *n* classes from the test set and remove them from the training set to simulate concept drift. The positive samples are the samples from dropped classes in the training set but remain in the testing set. The remaining samples in the training set from the known class are treated as negative samples. In our experiments we set *n* to [1, 2, 3, 4] values. We measure the precision, recall and, F1 score of the drift classifier with respect to classes dropped from the training set. Precision is measured by the percentage of the detected positive instances that were correct, recall is the proportion of correctly identified positive instances, and F1 score is the harmonic mean of recall and precision.

For testing how well the proposed method discovers new concepts we follow a similar testing protocol for evaluating incremental learning [39]. The classes are arranged in a fixed size and in random order. The network is trained in a class-incremental way on the available data and evaluated on the full test set. The *average incremental accuracy* [40] is the average accuracy of only those classes that have already been trained are used for evaluation. Finally, we measure the accuracy drop of the drift classifier with respect to the number of poisoned (injected) samples and concepts to verify the robustness towards adversarial drifts. We inject adversarial instances by corrupting a ratio $\mathbf{L}_{inst} \in [0.05, 0.10, 0.15, 0.20, 0.25]$ of selected instances and adversarial concepts are injected by generating a number $\mathbf{L}_{conc} \in \{1, 2, 3, 4, 5\}$ of small artificial concepts of $n = 250$ instances where $\mathbf{L}_{conc}$ are the number of concepts (classes) injected in the experiment.

*Training.* For each multi-class dataset, the train-test sets are in the ratio of 75% and 25%. Models are implemented with Pytorch. Adam [42] is used for the optimization. We used a learning rate of 0.0001 and 300 epochs for each dataset. When calculating the contrastive loss, we chose $\tau = 0.1$ as in [43]. We set $\lambda_1$ to 0.1 in the distance function for training the NCM classifier. The batch size for URL and Network are 128 and 256. For embedding networks we train a MLP with the architecture of 72-64-32, 134-64-32 respectively for Network and URL datasets. The threshold parameter $T$, which is used for identifying the drifted concepts into the new class is set to 3.5 and may vary depending on the choice of initial training data. The dimension of embedding network *h* is set to 32.

*Baseline Models.* We include 4 baselines for network data in our experiments - (1) Plain Autoencoder [44] (AE) trained *without* contrastive loss with the same number of layers and output dimensionality and use the prototypes to train NCM classifier to detect drift; (2) Adversarial detection and correction autoencoder network(ADCA) [45] combined with two-sample test *maximum mean discrepancy* MMD [46] with permutation test to obtain p-values. We use Alibi Detect [47] drift detection functionality to run baseline experiments. The autoencoder is trained on a model-dependent loss function designed to match the prediction probability distributions of the original and reconstructed instances, giving an adversarial score to test time samples. Tied with MMD, it can identify both adversarial examples and drifted samples in a single pass. (3) Incremental training iNtrusion System Over tiMe-stamped Network traffIc dAta (INSOMNIA) [16], a semi-supervised method that leverages incremental learning to detect drifts and transfer learning to discover new concepts. For Discovering new concepts they leverage the uncertainty sampling query technique, to select the data points that have the least certain prediction from the classifier in the previous batch for labeling either via active learning or NCM classifier to create new classes and retrain the classifier. In our experiments, the classifier was trained to support multi-class objective and we set the top $\sigma$, most uncertain traces to 50%; (4) Contrastive Autoencoder for Drifting detection and Explanation (CADE) [15] focuses on explaining drift, using contrastive autoencoder and a distance-based explanation method it performs drift detection. In our experiments, we keep the hyperparameters the same as the original proposal. For URL dataset we use AE and ADCA for baseline comparison. We also compare our proposed method with state of art Out of Distribution (OOD) methods based on (a) Euclidean distance metrics to score outlier samples, (b) One-class SVM [50] which employs maximum margin decision boundary to identify out of distribution samples and, (c) OpenMax [49] which applies Extreme Value Theory (EVT) [48] to derive sample weighting function and in turn use this function to reject OOD samples. In our experiments, each OOD method acts as the binary detector for in-out samples, where in samples are classes already in the training dataset and out samples are new classes not in training data and report Area Under the ROC Curve (AUROC) metric to assess OOD detection performance. We use the radial basis function kernel in our experiments to train OCSVM. The $v$ parameter is set to the expected anomaly proportion in the dataset, which is assumed to be known, whereas the $\gamma$ parameter is set to $1/m$ where $m$ is the number of input features. For euclidean distance measurement, we calculate the pair-wise distance of the incoming sample with the training data samples to score the outliers. We re-implemented the authors' published algorithm using the PyTorch framework [51] and used it in our experiments.

Finally, we compare our method with Multi-Layer Percepton (MLP) Classifier[52] to reflect non-drift and non-adversarial settings.

Table 1: Drift Detection results for Network and URL datasets. 4 baselines models – ADCA, AE, CADE, and INSOMNIA are compared for network datasets, and AE and ADCA are used for URL datasets. N classes are dropped from the training, to test concept drift detector accuracy. All measures reported( mean) over 5 runs)

| Models for URL Data | Precision | Recall | F1 Score |
|---|---|---|---|
| Number of Classes Dropped 2 | | | |
| Proposed Method | 0.928 | 0.885 | 0.906 |
| ADCA | 0.883 | 0.874 | 0.878 |
| AE | 0.736 | 0.707 | 0.721 |
| Number of Classes Dropped 4 | | | |
| Proposed Method | 0.831 | 0.853 | 0.842 |
| ADCA | 0.663 | 0.686 | 0.674 |
| AE | 0.612 | 0.601 | 0.606 |

| Models for Network Data | Precision | Recall | F1 Score |
|---|---|---|---|
| Number of Classes Dropped 2 | | | |
| Proposed Method | 0.965 | 0.921 | 0.942 |
| CADE | 0.905 | 0.891 | 0.897 |
| INSOMNIA | 0.882 | 0.819 | 0.849 |
| ADCA | 0.815 | 0.782 | 0.798 |
| AE | 0.788 | 0.766 | 0.777 |
| Number of Classes Dropped 4 | | | |
| Proposed Method | 0.836 | 0.793 | 0.813 |
| CADE | 0.716 | 0.693 | 0.704 |
| INSOMNIA | 0.682 | 0.639 | 0.659 |
| ADCA | 0.693 | 0.65 | 0.671 |
| AE | 0.637 | 0.623 | 0.63 |

Table 2: AUROC of OOD methods

| Method | Dropped Classes 2 | | Dropped Classes 4 | |
|---|---|---|---|---|
| | Network | URL | Network | URL |
| Proposed Method | 89.8 | 92.8 | 88.5 | 84.6 |
| One-class SVM | 62.9 | 76.5 | 53.4 | 64.8 |
| OpenMax | 61.7 | 79.8 | 52.7 | 74.7 |
| Euclidean distance | 52.3 | 68.7 | 51.3 | 61.9 |

## 6. Results and Discussion

We compare the proposed drift detector performance with baseline models, state of art OOD methods and highlight our observations with error analysis. First, let's discuss the effect of precision, recall, and F1-score metrics when the number of unseen classes increases in the test set. Table 3 summarizes the results of the experiment. Plain Autoencoder scores drop as new classes emerge at test time supporting our initial assumption of contrastive loss improves the drift detection model. ADCA drift detection model employs KL-divergence loss and MMD two-sample test to check the adversarial score of the incoming sample. This method gives similar performance when compared to the proposed method. But, as the number of dropped classes increases our proposed method accuracy still holds whereas other detectors performance deteriorates. In non-adversarial drift settings, CADE performed similarly to the proposed method. This can be attributed to the similarity in contrastive loss function it was trained on to detect drifts. INSOMNIA adaptation of NCM classifier to discover neighbor classes with active learning paradigm, when exposed to new samples may be one of the reasons for less degradation in the accuracy. Our method not only outperforms baseline models but also gives high recall and good precision, which is an important metric in cybersecurity applications. We inspected manually the missed samples by each method, we observed that some class labels have overlapping features due to labeling errors. For example, take the case of the website categorization dataset, the categories such as *hacking* and *malicious sources* have a large overlap of features like hosting infrastructure, whois information, etc which we think is one of the reasons for sources of errors in the detection. Compared to OOD detection methods the proposed method performed well in identifying new class samples. The distance metric combined with contrasive loss can be attributed to performance gain. Table 2 summazises the AUROC of the OOD experiments.

Table 3: Accuracy measures of MLP and Proposed Method on IDS data set trained as binary classifiers in a non-drift and non-adversarial settings

| Dataset | Model | Precision | Recall | F1 Score |
|---|---|---|---|---|
| *Benign vs Infiltration* | MLP | 0.998 | 1 | 0.998 |
| | Proposed Method | 0.996 | 0.999 | 0.997 |

Most of the adversarial attacks proposed in literature assume the i.i.d nature of data with the knowledge of *all* classes beforehand. In a typical setting, one may use optimization techniques to generate adversarial samples to test the robustness of the underlying classifier. These assumptions may not hold in the real world where data is prone to concept drift. Our experiments design to test the robustness without making the above-mentioned assumptions to reflect real-world scenarios. To test the robustness we inject adversarial concepts and drifts into the test set. The aim of the adversary is to evade the underlying drift detector module. We observed that the proposed method is stable with respect to stronger attacks as 20% of instances in testing data per class are adversary-induced drifts. The baseline methods which performed on par with the proposed method in non-adversarial settings suffered a steep degradation in both concept-based and instance-based poisoning experiments. This validates our design decision to employ both euclidean and geometric distances into the NCM classifier and making the class boundaries tighter. Figure 7 illustrates the accuracy drops of each model under adversarial drifts. Finally, we compare our method with the MLP classifier which does not take into temporal drifts i.e. we train both the models on the entire dataset and measure the accuracy. We observe that in this setting, our model performed on par with the MLP classifier. Table 3 summarizes the corresponding metrics.

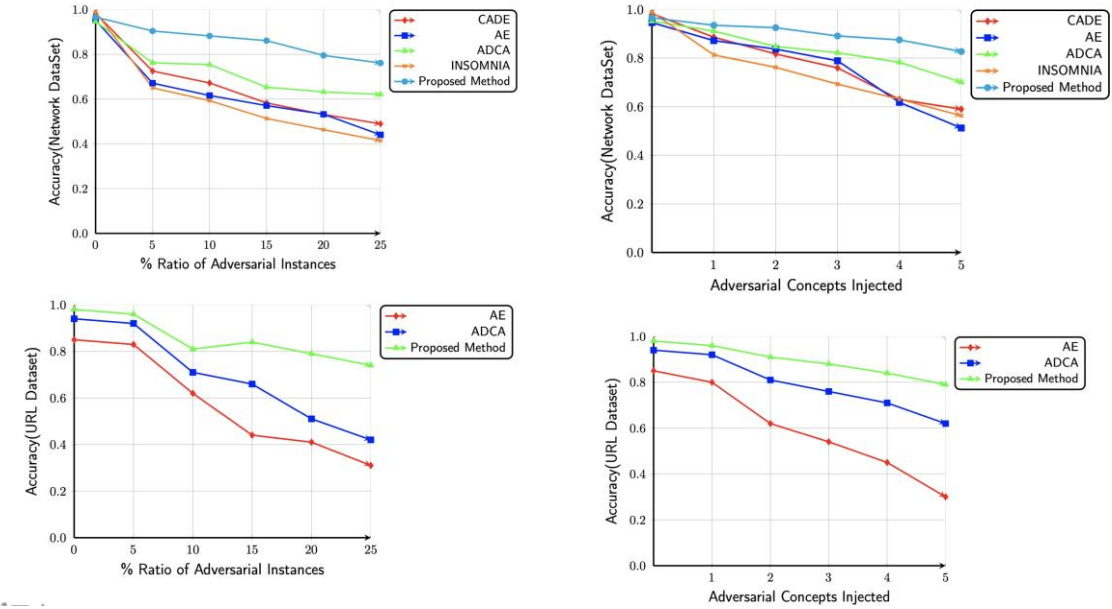

Figure 7: Accuracy drops of URL and Network drift models when adversarial concepts and instances are injected into training data

## 7. Related work

Our work combines several research areas, and we briefly discuss related work from these areas below.

**Drift detection for cybersecurity use cases:** Few works in literature have addressed drift detection problems in the context of cybersecurity [53]. In the field of malware detection, multiple works have addressed the problem of concept drift by treating malware data in an online fashion. Masud et.al. [54] trained an ensemble of classifiers with sequential data to address distributional drifts. Some works proposed the tracking of the changes in features of malware [55] addressed concept drift. Other works compared instance-wise comparison of training samples with new incoming samples to evaluate quality [17] and confidence of incoming samples and in turn identify drifts.

In their seminal work [14] authors proposed Tesseract framework to avoid experimental bias while measuring the model performance degradation over time due to concept drift. Metrics such as quarantine cost and labelling cost are baked into the workflow to provide reliable results in real-world settings. However, their approach does not account for adversarial drift which is the main focus of our study. The re-training/labelling strategy using either uncertainty samples(low confidence outputs) may also incur significant cost of human efforts.

Network intrusion detection use cases are treated as streaming settings [56, 57, 58] and a large body of work is already proposed in literature to address concept drifts in a streaming setting. In supervised setting, DDM (Drift Detection Method[59]), EDDM (Early Drift Detection Method[60]) and ADWIN (ADaptive WINdowing[61]) are popular approaches. Statistical tests over a given time window to test for errors or changes in trends is one the most used approach in an unsupervised setting. We refer readers to recent survey [2] for a thorough treatment of the subject. Our proposed setting can be used in a streaming fashion with no changes to the design. Very recent work[15] similar to ours leverages contrasive loss to detect concept drifts and explain the drift intensity using a distance function but they do not test the robustness of the methods in the presence of attacker induced drifts. All these works tackle the drift detection in a non-adversarial setting. In our work we aim to fill this gap by creating a robust detector and testing methods to improve robustness of the underlying drift detectors.

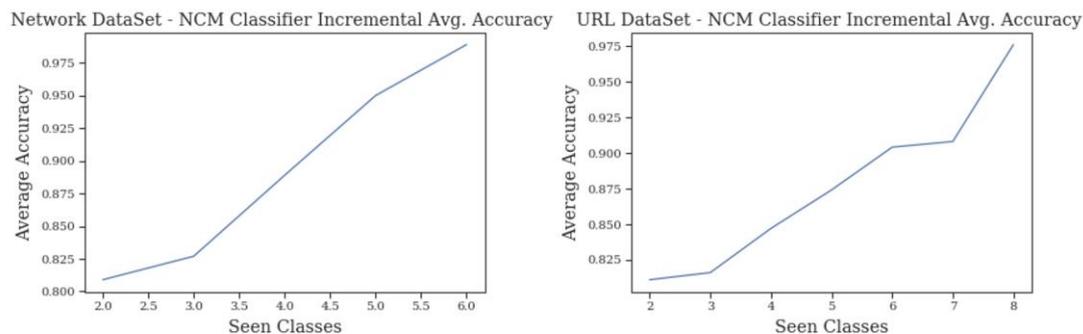

Figure 8: Incremental average accuracy gain of Drift classifier as more classes are exposed in incremental fashion

**Out of Distribution Detection(OOD)**: Another related line of work is out-of-distribution sample detection. At a high level, OOD detectors compare where the new sample falls in-out of distribution based on a distance, density, or distribution metric: (a) Learning-based outlier detection methods [50] leverage ML or statistical models to learn the distribution of clean data and detect out-of-distribution samples (including adversarially corrupted samples); (b) Error-detection methods [62] rely on a combination of the logic-rule and ML models to identify corrupted samples; For detecting outlier for high dimension data multiple methods are proposed in literature – Deep Outlier Exposure[63], Ensemble of Leave-out Classifiers [64], Deep Mahalanobis Detector [65] for calculating the class-conditional outlier score from the deep features, and OpenMax[49], which uses mean activation vectors of classes observed during training followed by Weibull fitting to determine if a given sample is novel or out-of-distribution. These works are relevant but they are solving for a different objective. We compared our method with some of the state of art OOD methods and show that our method performs better.

**Contrasive Learning** The idea of pulling similar samples together and pushing dissimilar samples apart comes from exemplar learning [66]. Contrastive loss function has been adapted in the self-supervised task for learning representations without labels [26]. Models based on contrastive losses have significantly outperformed other approaches [67, 26]. Most similar to our work is [68] who learn cluster prototypes via contrastive loss, but instead of creating prototypes at training time, we learn class-boundaries in an incremental fashion.

## 8. Limitations

We want to highlight some limitations of our method. In our experiments, we have not thoroughly studied the time, and space complexity of the algorithm. In real world settings, production systems using drift detectors are designed to work in-line with the deployed classifier. The inference times and response times are critical for any production grade system. In our experiments, we identified many misses by the drift detector that are mainly because of labeling errors or large feature set overlap. In real-world setting, training set is prone to labeling errors [69] and may lead to false sense of drifts, we plan to tackle this problem in future. Threat models and attacks on drift detectors are rare, as this is auxiliary classifier and acts as a sub-module to main classifier, adversary has less visibility into the internal working of the drift classifier. In this work as part of testing process we devised two novel attacks for drift detectors. As a future work we aim to study in depth these attacks under different threat models. In security sensitive systems, users expect a through and detail explanation of model outputs [71, 70, 72] with decision contexts, our system does not provide an explanation interface but we envision that with the querying k-nearest neighbors samples of drifted samples in latent space and extracting the features, we can support explanation interfaces and we defer this work for future.

## 9. Conclusion

In this paper, we propose a novel drift detection method to address the concept drift in real-world security applications. Using contrastive loss objective and NCM classifier we train a robust drift detector that is immune to attacker-induced drifts and effective in identifying drifted samples. We also devise a recursive update mechanism to discover new classes from drifted samples and retrain the detector in an automated fashion. Two security-relevant datasets were used to evaluate the proposed method and show that it is effective in the drift detection and identifying new classes, which are previously unseen in the training set.